\documentclass[aps,prx,twocolumn,amsmath,amssymb]{revtex4}
\usepackage{graphicx}
\usepackage[colorlinks,citecolor=blue,linkcolor=magenta]{hyperref}
\usepackage[usenames,dvipsnames,svgnames]{xcolor}
\usepackage{txfonts}

\begin{document}
\title{Yu-Shiba-Rusinov states of single magnetic molecule in an $s-wave$ superconductor }

\author{Saurabh Pradhan}
\author{Jonas Fransson}
\affiliation{Department of Physics and Astronomy, Box 516, 75120, Uppsala University, Uppsala, Sweden}

\begin{abstract}
    We use the numerical renormalization group theory to investigate the Yu-Shiba-Rusinov (YSR) bound state
    properties  of single magnetic molecules  placed in an s-wave superconducting
    substrate. The molecule  consist of a large core spin and a single orbital, 
    coupled via an exchange
    interaction. The critical Coulomb  interaction for the singlet/doublet transition decreases
    in the presence of this exchange interaction for both ferro and anti-ferromagnetic couplings.
    The number of YSR states also increase to two pairs, however, in the singlet phase, one of the pairs have
    zero spectral weight. We explore the evolution of the in-gap states using the 
    Anderson model. Away from the particle-hole symmetry point, the results suggest a doublet-singlet-doublet transition as
    the on-site energy is lowered while keeping the Coulomb interaction fixed. To understand these results, we  write down an
    effective model for the molecule in  the  limit of large superconducting order parameter. Qualitatively,
    it explains the various phase transitions and spectral nature of the in-gap states. Finally, we analyze the
    effects of magnetic anisotropic fields of the core spin on in-gap states. Due to internal degrees of freedom of the spin
    excited states, a multitude of new states emerges within the gap. Depending on the sign and strength of the uniaxial
    anisotropic field, the results indicate up to three pairs of YSR states.
\end{abstract}
\maketitle

\section{Introduction}
Nanoscale devices coupled in a tunnel junction give unique opportunities to study quantum
many-body effect of impurity systems. Recent years, there have been tremendous advancements
over the control of such devices, where the number of  electron is controlled electrostatically in a
small restricted region. Whenever these devices contain an odd number of electrons, the Kondo
effect\cite{scott,hewsonBook,colemanBook}  arises due to multiple spin-flip scattering processes.
Quantum dots\cite{dots1,dots2},
magnetic addatoms\cite{addatoms1,addatoms2,addatoms3}, and magnetic molecules\cite{molecule1,
molecule2,molecule3}
are examples of microscopic systems which display the Kondo effect when mounted
in a junction between metallic electrodes.

Low temperature experiments with
localized magnetic moments adsorbed onto superconducting surface display the emergence of
bound states inside the superconducting gap. 
This was first measured 
using scanning tunneling microscopy and spectroscopy\cite{exp1} for  single magnetic impurities,
which has subsequently been reproduced by various groups under different
experimental conditions, such as, magnetic field\cite{exp2,exp3}. 
Theoretically, the emergence of in-gap bound states was predicted
by Yu\cite{Yu}, Shiba\cite{Shiba}, and Rusinov\cite{Rusinov},
using semi-classical approaches where, specifically, the spin moment was treated as classical.
Quantum effects of magnetic impurities were later studied within mean-field
theory\cite{mft1,mft2}, perturbation theory\cite{purtb1,purtb2}, and
numerical renormalization group (NRG) theory\cite{nrgsc1,nrgsc2,nrgsc3}.

The ground state of a quantum magnetic impurity in a metal substrate is a Kondo singlet with a characteristic
energy scale related to the Kondo temperature ($T_k$). 
However, in a superconductor, the substrate
electrons form Cooper pairs which are not compatible with the Kondo singlet state. The fundamental interactions associated with the superconducting
gap ($\Delta_\text{sc}$) and the Kondo temperature compete with each other, and at large $\Delta_\text{sc}$ the
ground state becomes a doublet formed by substrate and impurity electrons. The ratio of the two
energy scales determines the nature of the ground state and energies of the bound states inside the gap.
The bound state energy coincides with the energy of the edge of the superconducting gap for weak
coupling between the spin moment and the surface states, and move inside the gap for increasing
coupling strength, eventually crossing the Fermi energy when the two energy scales are similar
$T_k \approx \Delta_\text{sc}$.

The bound states always come as in pairs of particle-hole symmetric states
around the Fermi energy. Recently, more than one pair of Yu-Shiba-Rusinov (YSR) states
were observed\cite{orbital2,orbital3} in experiments with magnetic molecules. Multiple pair of
YSR states have been attributed to the presence of many orbitals in the molecule. The coupling of
these orbitals with the substrate is not uniform due to the different nature of the orbitals.
As a result, the energies of the YSR states of the different orbitals may have different energies and weights.
Results from NRG calculations of large spin moments with magnetic anisotropy, show that multiple pairs of YSR
states may appear\cite{orbital4} due to internal spin excitation. These
calculations do, however, not take into account differences due to the localized and delocalized nature of
the $d$- and the ligand orbitals, respectively.

Here, we have considered an Anderson impurity model, in which superconducting surface
states play the role of the reservoir electrons, coupled with a core spin. Due to the
presence of the core spin, multiple YSR states emerge inside the superconducting gap.
This model naturally  reflects the geometry
of the large spin molecular systems such as Fe$_8$\cite{Fe8}, Mn$_{12}$\cite{Mn12}
and  transition metal phthalocyanines\cite{orbital2, frankeReview}.
The ligand orbitals of
the molecule form degenerate orbitals which couple with the surface electrons
in the substrate. For simplicity, we have considered a single orbital Anderson impurity model for
the ligand orbitals. The $d$-orbital electrons of the transition metal atom
do not hybridize much and, therefore, form a local magnetic moment. We have assumed
that the spin moment of the transition metal atom interacts only with electrons in the
ligand orbital via exchange and has no interaction with the substrate electrons. We have considered the magnetic moment of the
transition metal atom to be large ($S >1/2$). Due to
spin-orbit coupling and spatial structure of the substrate, we have also
included magnetic anisotropy for the core spin. In case of phthalocyanine
molecules, some $d$-orbitals can form a core-spin without hybridizing
with the substrate while other $d$-orbitals do hybridize. The latter
scenario do give rise to a pronounced Kondo effect\cite{orbital5}.

This paper is organized as follows. In Sec. II, the model is defined
and we present a brief description of the NRG method. In Sec. III, we derive an effective model
for our system in the limit of a large superconducting gap. This gives a better
understanding of the NRG results and qualitatively explains various properties
of the YSR states. In Sec. IV,  we present the NRG results. In Sec. IV A, we
discuss the proximity induced superconducting order parameter in the molecule,
whereas in Sec. IV B, we discuss the behavior of the YSR state as a function
of the Coulomb interaction for different values of the exchange interaction.
In Sec. IV C, we discuss the spectral weight of the YSR 
states. Single particle transition from the ground state is only possible when the 
spectral weight is non-zero.
In Sec. IV D, we change the on-site energy of the orbital away from the particle-hole
symmetric point,
while in Sec. IV E, finally, we look at the behavior of the YSR as we turn on the
magnetic anisotropy field.
The paper is concluded and summarized in Sec. V.

\section{Model and Methods}
\begin{figure}[t]
    \centering
    \includegraphics[width=0.49\columnwidth]{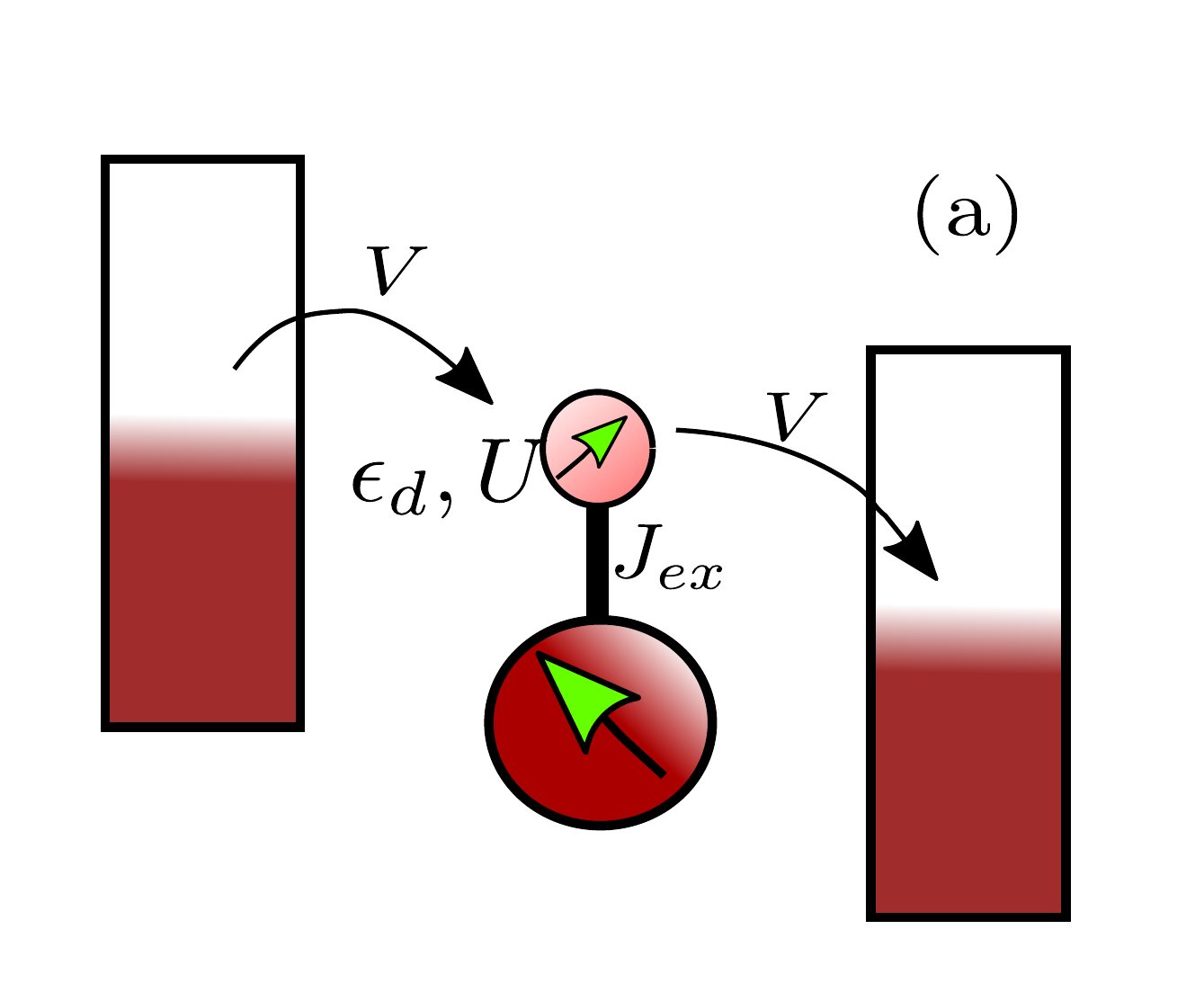}
    \includegraphics[width=0.49\columnwidth]{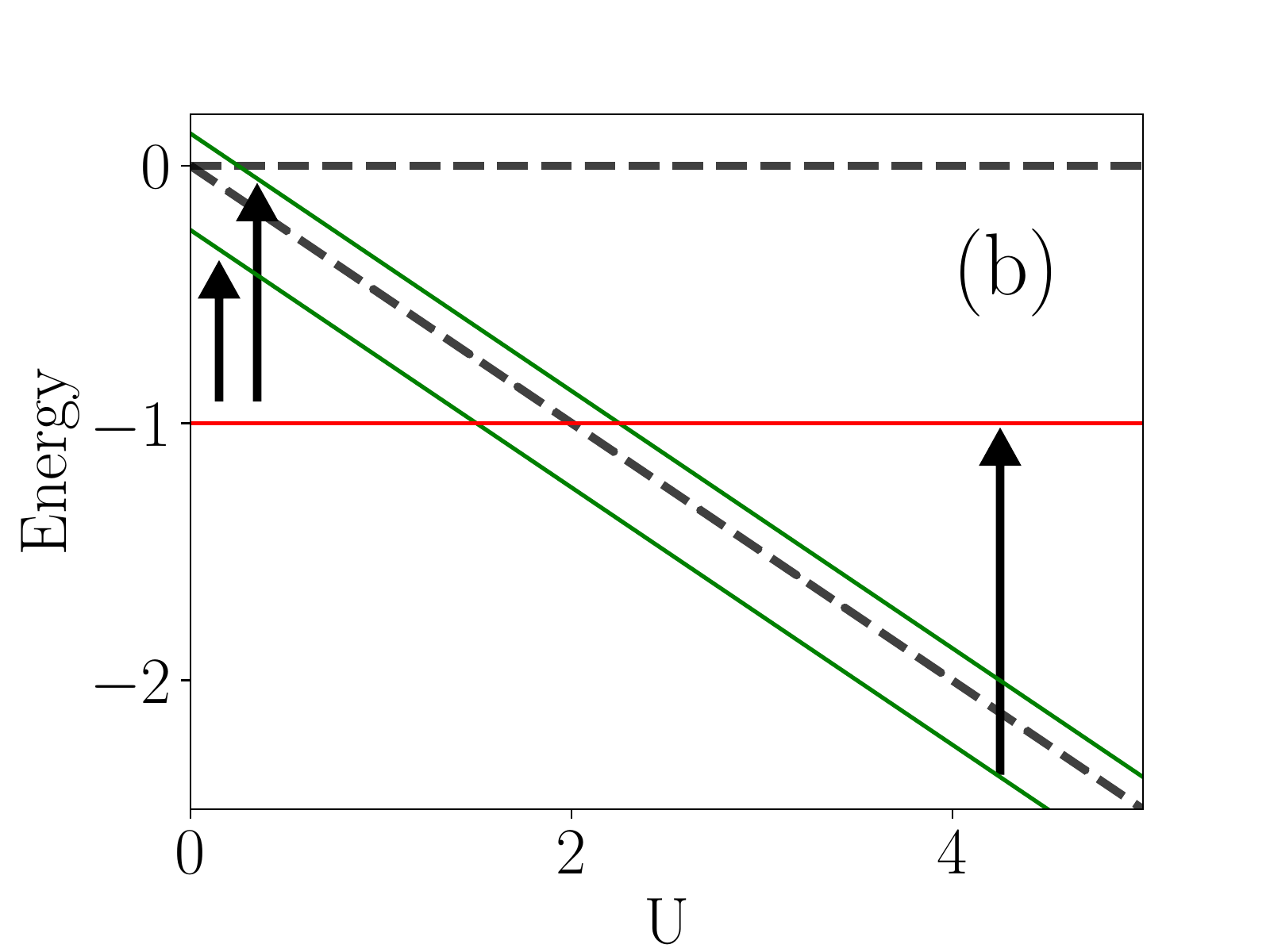}
    \caption{(a) Schematic diagram of  the magnetic molecule in a
        tunnel junction with finite bias voltage ($V$). (b) Energy
        eigenvalues of the effective Hamiltonian as function of Coulomb
        energy ($U$) for the molecule in the large $\Delta$ limit. Solid
        lines are for eigenvalues when $J_\text{ex}$ and $\Delta$ are
        finite, whereas dashed line represents vanishing values, that
    is, $J_\text{ex}=\Delta=0$.}\label{fig:schematic}
\end{figure}

We consider the magnetic molecule to be embedded in a tunnel
junction between metallic electrodes, as depicted in Fig.\ref{fig:schematic} (a). The total Hamiltonian of the system is given by
\begin{subequations}\label{eqn:hal}
        \begin{align}
        H =& H_\text{mol} + H_\text{sc} + H_\text{T}
        \label{eqn:hal1}\\
        H_\text{mol} =& \epsilon_d n_d + U n_{\uparrow}n_{\downarrow} +   J_\text{ex}\mathbf{s}\cdot \mathbf{S}  + H_\text{S}
        \\
        H_\text{S} =& D S_z^2
        \\
        H_\text{sc} =& \sum_{k\sigma}
        \epsilon_k c^\dag_{k\sigma}c_{k\sigma} - \Delta_\text{sc}\sum_{k}  (c^\dag_{k\uparrow}c_{-k\downarrow} + H.c.)
        \\
        H_\text{T} =& \sum_{k\sigma}V_{k}(c^\dag_{k\sigma}d_{\sigma} + h.c )
    \end{align}
\end{subequations}
The molecule consists of a single orbital labelled with the on-site energy $\epsilon_d$ and Coulomb repulsion energy $U$, $n_\sigma = d_\sigma^\dag d_\sigma$ is the number operator for each spin, $d_\sigma^\dag$ is the creation operator for the orbital, and $n_d = n_{\uparrow}+n_{\downarrow}$ is the electron occupation number of the orbital.
The orbital spin ($\mathbf{s}$) and the core spin ($\mathbf{S}$) interact via exchange interaction ($J_\text{ex}$).
To describe magnetic molecules in general,
we have included an
anisotropy field ($D$) for the core spin.
For simplicity,  here we have not included  transverse anisotropic term $E(S_x^2-S_y^2)$. 
The electrons in the superconducting substrate are described by $s$-wave Bardeen-Schreiffer-Cooper (BCS) mean-field Hamiltonian
($H_\text{sc}$).
The first term of $H_\text{sc}$ describe the kinetic energy part of the free electrons
in the substrate. In the absence of superconductivity, the substrate
has a constant density of electron states, $\rho_0 = 1/2\cal{D} $, within $[-\cal{D},\cal{D}]$ with a bandwidth $2\cal{D}$.
Henceforth, we take $\cal{D}$ as the absolute energy scale of the system and set it to be $\cal{D} = $1. We have added a BCS order parameter  $\Delta_\text{sc}$, and the
temperature dependence of the $\Delta_\text{sc}$ is neglected, as we only discuss the ground state  properties. Here, we
have neglected electron-electron interactions in the substrate.
We also fix $\Delta_\text{sc}/{\cal D} = 2\cdot10^{-4}$ and $S = 2$.

In this work  we use NRG theory\cite{nrg1,nrg2,nrg3,nrg4,nrg5,nrg6,nrg7},
which is an unbiased non-perturbative method that works perfectly at both
zero and finite temperatures. First, we 
discretize the non-interacting  substrate electrons such that the electrons are described by a finite number of
states, logarithmically separated from each other. Second, we transform this system into a linear chain which begins with the  molecule. The on-site energy of the linear chain
is put to zero for our constant density of electron states bath electron
and the hopping elements decrease exponentially with increasing distance away from
the molecule. We use the NRG
discretization parameter $\Lambda = 2.5$ throughout this paper.

The exponential decrease of the energy scale of successive sites in the linear chain
ensures the success of the NRG method for the metallic substrate. For
the superconducting substrate it was initially thought, however, that the superconducting gap  $\Delta_\text{sc}$ would cause
a problem for the NRG iteration, since the argument
of energy scale separation  no longer holds for large Wilson chains.
In other words, the perturbation of adding a site to the  Wilson chain
is no longer sufficiently small to allow truncating the NRG iteration at site $N$ where $\Lambda^{-N/2}$ is comparable
to $\Delta_\text{sc}$\cite{nrg1,nrg2}. It was, nevertheless, pointed out
that the energy scale separation of NRG works
even beyond this value of $N$, and that the perturbations become even
smaller with finite  $\Delta_\text{sc}$\cite{nrg5}.
Hence, in the presence of superconductivity, the NRG approximation becomes even more accurate.

\section{Large $\Delta$ limit}
Before discussing the NRG results we consider a simplified version of the model in Eq. (\ref{eqn:hal}), obtained in  the
limit of large $\Delta $, in order to gain some understanding of the expected behavior of the many-body
YSR bound states. As is illustrated for the single orbital Anderson model  in Refs.\cite{Rozhkov,Tanaka}, the substrate induces superconducting order in the quantum dot when the superconducting
gap is the largest energy scale compared to any other energy scales of the system.
In this limit, the self-energy, due to the bath electrons, give only finite
off-diagonal component in the Bogoliubov-de Gennes basis for energies much smaller than
the superconducting gap. As a result, we can write an effective Hamiltonian for the
system. This procedure can also be applied to the molecular system and
the effective low energy Hamiltonian can be written as
\begin{align}
    H_\text{eff} = \epsilon_d n + U n_{\uparrow}n_{\downarrow} +
    \tilde{\Gamma}\big(d_{\uparrow}d_{\downarrow} + h.c \big) +
    J_\text{ex}\mathbf{s}\cdot \mathbf{S}
    ,
    \label{eqn:effhal}
\end{align}
where $\tilde{\Gamma}$ is the induced superconducting order in the molecule.

In the
absence of the exchange term between the orbital spin  and the core spin, the ground
state behavior changes as function of $U$ at $U/2 = \tilde{\Gamma}$.
The ground state is a doublet state (the anti-symmetric combination of $|0\rangle$
and $|\uparrow\downarrow\rangle$) for small $U$ for the symmetric
Anderson impurity model ($\epsilon_d = -U/2$), while for large $U$, the ground
state is a singly occupied the state. The transition between these two state occur
at $U = 2 \tilde{\Gamma}$. The ground state degeneracy also changes from one
to two electrons across this transition, resulting in that the expectation values of various operators change discontinuously.

The energy difference between the ground and first excited states
mimics the behavior of the YSR bound states when $\Delta_\text{sc}$
is comparable to the other energy scales in the system. Within this effective
model, we can see that the bound state energy first decreases with increasing $U$ and approaches zero at the transition point, while it increases
again after the transition point, see Fig.\ref{fig:schematic} (b).

In the presence of a finite exchange interaction between the orbital spin and
core spin, the doublet ground state remains unaffected while the degeneracy of the singly
occupied state is lifted.
The energy of the singlet states are now  $-U/2 +J_\text{ex}S/2$  and
$-U/2 -J_\text{ex}(S+1)/2$. In Fig.\ref{fig:schematic} (b), we have shown the energies of the effective
Hamiltonian as function of $U$. The
first excited state is split below a critical energy $U_c$, whereas the ground state splits at larger $U$. Even though there are more states within the gap, not all states are
visible in the single particle spectrum at zero
temperature. This is clear since for $U<U_c$ transitions between the 
single state to both doublet states are possible, while for $U>U_c$ 
only transitions from the double state with the lowest energy to 
the single state are allowed, given that the temperature is 
sufficiently low to prevent thermal excitations of the 
second double state. The arrows in Fig.\ref{fig:schematic} (b) indicate all possible transitions.
As a result, a single pair of YSR states emerges in the singlet phase and while two pairs of
YSR states should be observed in the doublet phase.
We also observe that $U_c$ is
 shifted towards the lower value for both positive
and negative values  of the exchange
interaction. This can be understood from the fact that the exchange
interaction always lowers the energy of the many-body singlet state.

\section{NRG Results}
\subsection{Induced Superconductivty}

\begin{figure}[t]
    \includegraphics[width=\columnwidth]{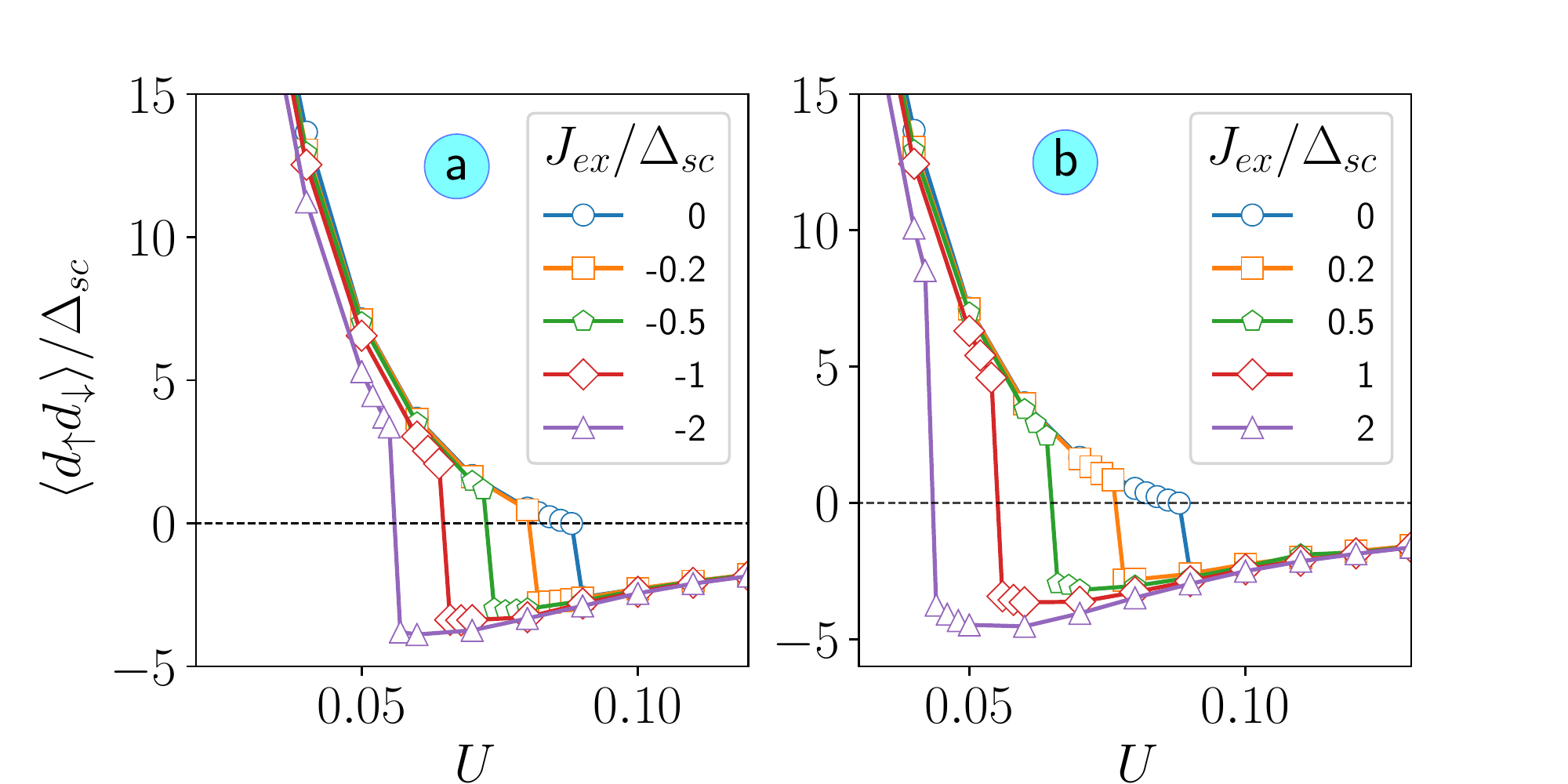}
    \caption{The expectation value of superconducting order
        of the orbital label  at zero temperature for different
        values of Coloumb Interaction at fixed values of the exchange
    interaction $J_\text{ex}/\Delta_\text{sc} \leq 0$ (a), $J_\text{ex}/\Delta_\text{sc} \geq 0$ (b).}\label{fig:scOrder}
\end{figure}

Due to proximity, the superconductor induces a finite pairing potential, or, 
supconducting order parameter in the molecular the dot. In Fig.\ref{fig:scOrder}
we show the expectation values of the induced order parameter as function of the 
Coulomb interaction. First, consider the case of vanishing exchange interaction,
$J_\text{ex}=0$ (blue). It can be clearly seen that the largest values of the order 
parameter is reached in the non-interacting limit, $U\rightarrow0$, and decreases 
with increasing interaction strength $U$. This can be explained as an effect of 
that the doubly occupied (and empty state for the symmetric Anderson model) state 
is less favorable for large $U$. Hence, since the BCS states are a combination of 
the empty and doubly occupied states, the induced superconducting order becomes 
suppressed as the Coulomb interactions become increasingly influential.

A further increase of the Coulomg interaction strength, $U$, leads to that the superconducting order parameter changes sign at the critical interaction energy $U_c$. At this transition point, the ground state of the molecule changes from doublet ($U<U_c$) to a singlet ($U>U_c$) state. The discontinuity of the induced order parameter at $U=U_c$ reflects the change in the  degeneracy of the ground state when it  undergoes a transition from doublet to singlet state. In the context of Josephson junction, this discontinuity is
related to the \emph{so-called} $0-\pi$ transition. One important thing to notice is that for Coulomb interactions weaker than the critical energy $U_c$, the induced superconducting order has the same phase as the substrate, whereas it is phase shifted by $\pi$ in the large $U$ limit.
While the order parameter remains negative for $U>U_c$, its magnitude decreases.

\begin{figure*}[t]
    \includegraphics[width=\textwidth]{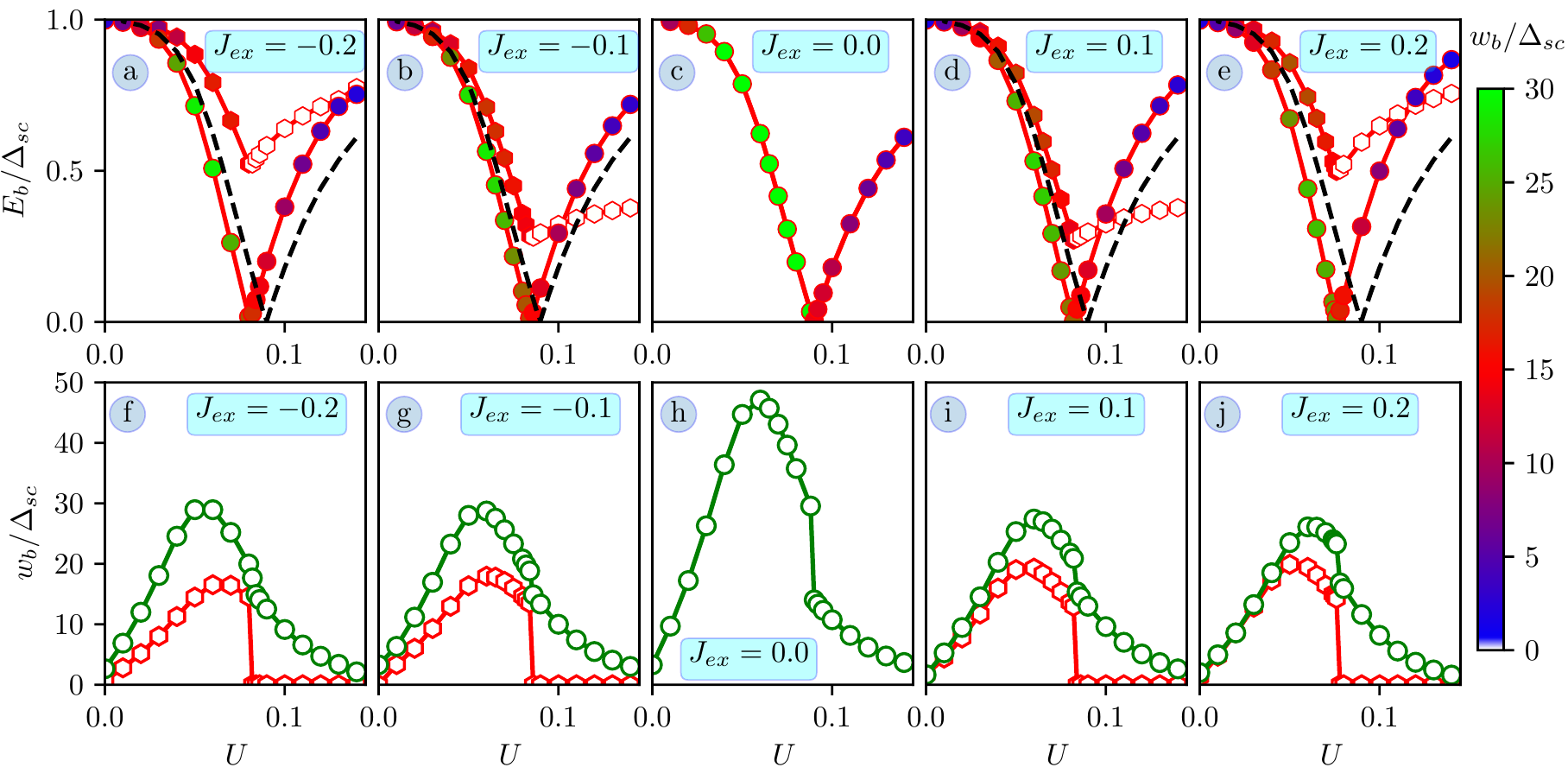}
    \caption{Bound state energies (a-e)  and their spectral weight (f-j)  of the in  gap states  for different
        value of Coulomb Interaction  at fixed  exchange interaction $J_\text{ex}/\Delta_\text{sc}$ = -0.5 (a,f),
        -0.2 (b,g), 0 (c,h), 0.2 (d,i), 0.5 (e,j). The spectral weight of each states are also color coded. The
        dashed line indicate the YSR states for $J_\text{ex}=0$.
    }\label{fig:specJex}
\end{figure*}

Next, we include a finite exchange interaction, $J_\text{ex}\neq0$. Previously, in 
the bound state analysis of the effective model, Eq. (\ref{eqn:effhal}), it was 
shown that the critical energy $U_c$ decreases in the presence of a finite exchange
interaction, irrespective of being ferro and anti-ferromagnetic. Here, our numerical 
results corroborate this conclusion, which can be seen in Fig.\ref{fig:scOrder}. 
Specifically, the sign change of the induced superconducting order parameter shifts 
to a lower $U$ for increasing $|J_\text{ex}|$. The value of $U_c$ reduces much faster in the case
of positive exchange interaction compared to negative value of the exchange interaction.
The core spin ($S$) and the orbital spin ($s=1/2$) have eigenstates that can be categorized as a triplet and a singlet, 
with corresponding energies $ J_\text{ex}S/2$  and $-J_\text{ex}(S+1)/2$, respectively, in the absence of the substrate.
Form this it is evident that the ground state energy decreases  faster when $J_\text{ex}$
is positive. This makes the reduction in $U_c$ is larger when $J_\text{ex} > 0 $.
Near the transition point, 
$U\simeq U_c$, the induced superconducting order parameter is strongly 
renormalized by the finite exchange interaction. The absolute value of the 
order parameter is reduced comared to the case when $J_\text{ex}$ is zero.

\subsection{Bound states}
The dimensions of the Hilbert space of the Wilson chain increases by a factor of 4 for each added site. Therefore, we
discard higher energy states after a few iterations to keep the size of the Hamiltonian manageable. The maximum number of
states that we retain is of the order of 5000. After sufficient number
of NRG iterations ($N=40$), we obtain all the pertinent many-body states. This includes a  
continuum  of  states above the superconducting  gap and few states below. The states within
the superconducting gap are known as Andreev\cite{andreev}, YSR states\cite{Yu,Shiba,Rusinov}, and we shall use the latter 
nomenclature for the remainder of this article.
In Fig.\ref{fig:specJex} (a) - (e), the YSR state are plotted as function of the 
Coulomb energy $U$ of the molecule level, for various values of the exchange interaction $J_\text{ex}$.
Here, the on-site energy of the molecule is $\epsilon_d = -U/2$; symmetric Anderson model.
In this case, the single particle spectrum of the  system is particle-hole symmetric.
There exists a negative energy YSR state for each positive energy YSR
state; not shown in this figure.
By turning off the exchange interaction, that is, putting $J_\text{ex}=0$, the model 
is reduced to an Anderson model with a superconducting reservoir. The associated positive 
energy YSR state for this case is shown in Fig.\ref{fig:specJex} (c).
It can be seen that there is only one bound state, as expected under degenerate 
conditions, emerging near the edge of the superconducting gap for small values of $U$. 
The doublet state   is  ground state  and singlet state becomes the first excited state
for lower Coulomb energy. With increasing $U$, the bound state shifts into the gap, 
coinciding with the Fermi level ($E_b-E_F=0$) at the critical energy $U = U_c$. The 
ground state of the molecule changes from doublet to singlet state at this critical 
energy ($U_c$). For Coulomb repulsion energies $U>U_c$, the bound state energy shifts 
away from the Fermi level.
At larger $U$, singlet state is the ground state and doublet state is first excited state.
In the metallic system, Kondo temperature is only  energy scale of the system and 
the ground state is Kondo singlet state. Whereas in the presence of the 
superconductivity, $\Delta_\text{sc}$ becomes a second energy scale of
the system. The BCS state of the superconductor and Kondo singlet state 
compete  and giving rise
to a singlet to doublet transition.

In our previous discussion of the effective model, Eq. (\ref{eqn:effhal}), we argued
that a large superconducting order parameter ($\Delta_\text{sc}\rightarrow\infty$) 
accompanied by a finite exchange interaction ($J_\text{ex}$) splits the singly 
occupied states. The result is also valid for finite $\Delta_\text{sc}$, however,
slightly modified to involve a lifting of the $2S+1$-fold degeneracy of the Kondo 
singlet. The $2S+1$-fold degeneracy originates from the core
spin ($\mathbf{S}$) in the absence of the interaction between the core spin and
the molecular orbital. Due to internal spin excitations in presence of the 
exchange iteration, new states emerge within the superconducting gap, as 
shown in Fig.\ref{fig:specJex}. Due to finite exchange interaction, there are 
three low energy states arises (one doublet state and two singlet states).  
At zero temperature, one of these states becomes the ground state and two other state 
becomes higher energy excited state. These two higher energy states becomes the 
two YSR state that  appear in the positive part of  spectrum 
both for positive and negative values of $J_\text{ex}$. These two states arise from the transition
doublet state to singlet states, see Fig.\ref{fig:schematic} (b), or the other way around 
depending on the values $U$. One of the bound state energies coincides with the Fermi energy 
at a critical $U=U_c$. One important thing to notice here is that $U_c$ always
reduces both for ferromagnetic and anti-ferromagnetic exchange interaction. 
The energy of the second bound state remains finite for all  
values of $U$. The energy of this bound state first decreases with $U$,
has a minima at $U = U_c$, and increases again for increasing value of $U$. This 
state appears because of the transition from doublet state to higher energy
singlet state for $U<U_c$. For $U>U_c$, however, singlet to singlet transitions 
give rise to this state. The energy of this YSR state is indicated as filled and empty 
hexagon in Fig.\ref{fig:specJex} (a), (b), (d), (e), respectively.
The energy difference between two singlet states increases
with the absolute value of $J_\text{ex}$. As a result, the energy of this YSR state
(empty hexagon) steadily increases with $|J_\text{ex}|$  for $U>U_c$. 
 With  large enough $|J_\text{ex} |$ compared to $\Delta_\text{sc}$  this 
 bound state merges with the continuum states above the superconducting gap.

\subsection{Single particle spectrum }
Next, we make connection with experiments through a discussion of the single 
particle spectrum of the system. This can measured in the experiments, for 
instance, using scanning tunneling microscopy. The conductance thus measured 
is proportional to the local density of electron states of the single orbital. 
In the zero temperature limit, the local density of electron states contain 
two contributions, one from the states within the superconducting gap
and other from the states outside the superconducting gap. 
Formally the local density of  state can be written as 
$A(\omega)  = A_1(|\omega| < \Delta_\text{sc}) + A_2(|\omega| > \Delta_\text{sc}) , 
A_1(\omega)  = \sum_{i}w_b^i\delta(\omega - E_b^i) $. 
In this article we 
will  not  discuss the continuam part of spectrum ($A_2$) and we are only interested 
in the in-gap part of the spectrum ($A_1$). 
The spectral weights and corresponding energies of these in-gap states are 
plotted in Fig.\ref{fig:specJex}, for both positive and  negative values 
of the exchange interaction $J_\text{ex}$. Here, the on-site energy of the orbital
is $\epsilon_d=-U/2$; the symmetric Anderson model, imposing particle-hole
symmetry in the spectrum such that the bound states with positive
and negative energies have the equal weights. Previously, it has been shown that the spectral
weight is discontinuous at $U=U_c$\cite{nrgsc2}. Due to the presence of
the exchange interaction, two YSR states with finite weights arise in the spectrum
for $U<U_c$,  which is consistent with the large $\Delta_\text{sc}$ model
(see the arrows in Fig.\ref{fig:schematic} for smaller $U$). The two 
bound states arise from transitions between the doublet ground
state and the two singlet states. These two YSR states reduce to 
one when $J_\text{ex}$ is zero. It can be seen that both weights
increase with $U$, and that the weight of state with the higher energy
state drops to zero at $U=U_c$; indicated by empty symbols in
Fig.\ref{fig:specJex}. The weight  of the other state shows a 
discontinuity at $U=U_c$. The size of the jump at the discontinuity 
depends on the sign of the exchange interaction $J_\text{ex}$, which 
is more prominently appearing for anti-ferromagnetic exchange interaction.
To the right of the transition point, that is, $U>U_c$, the spectral 
weights gradually decrease with increasing $U$.

We summarize the discussion of the single particle spectrum by 
noticing, that below the critical point, $U<U_c$, two pairs of 
YSR states emerge in the superconducting gap whereas only one 
pair is observed above, $U>U_c$. This is one important 
result of this article. In the molecular set-up, it should 
be possible to vary the ratio $U/\Gamma$, and we expect that 
it through such variations should be possible to observe a 
transition between two pairs and one pairs of YSR states.

\subsection{Away from particle-hole symmetry}
\begin{figure}[t]
    \includegraphics[width=\columnwidth]{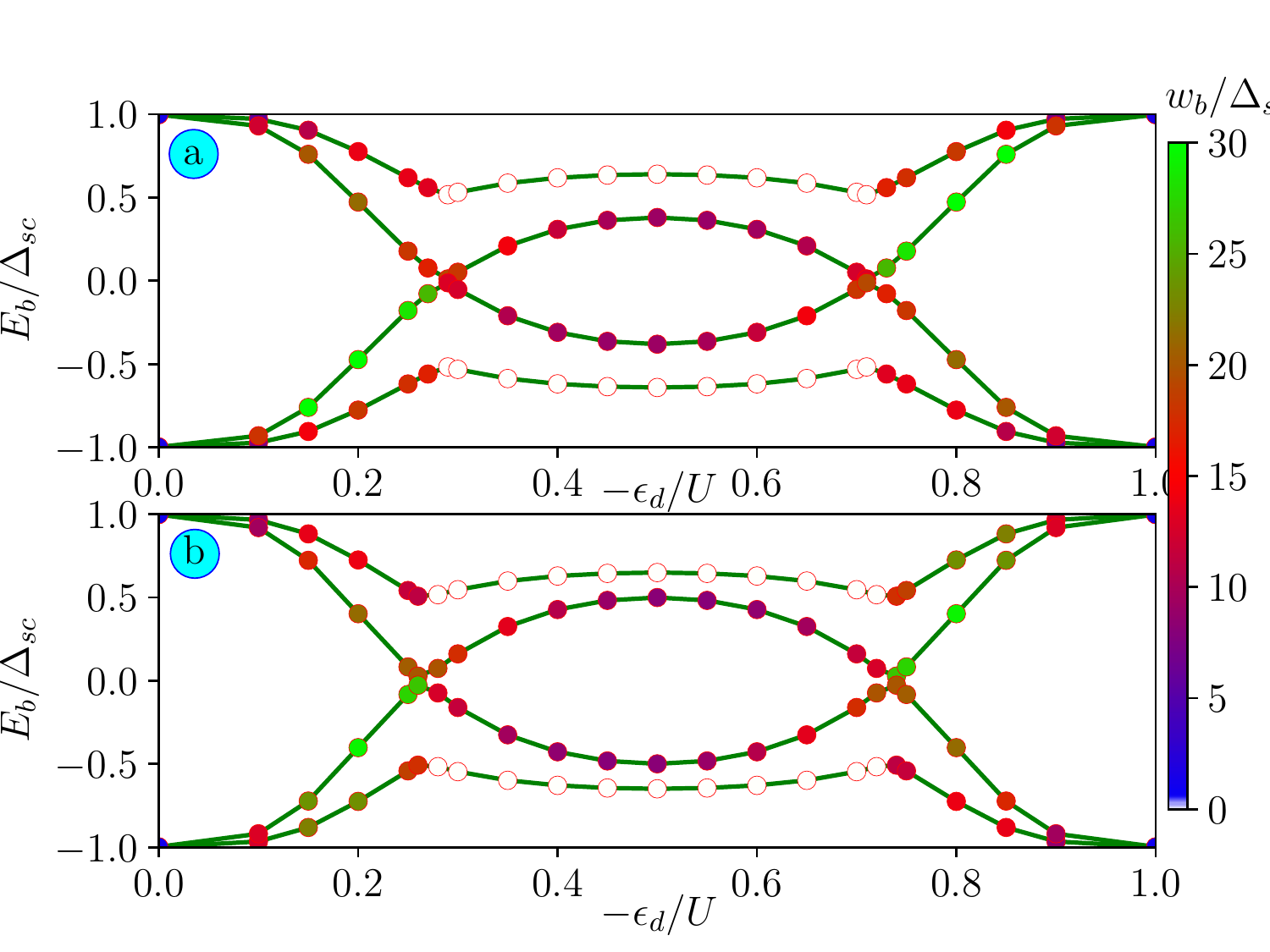}
    \caption{Bound state energies of the molecule  for different value of  on-site energy of the
        orbital away from the
    particle-hole symmetric point  for both ferromagnetic $J_\text{ex} = -0.2$ (a) and anti-ferromagnetic $J_\text{ex} = 0.2$ (b)  at a fixed $U=0.10$. }\label{fig:assymetric}
\end{figure}

Even though the symmetric Anderson model is most often used in the existing 
theoretical literature, such a symmetry is most likely not present in a 
realistic experimental set-up. Apart from this, it is also possible to
effectively shift the on-site energy by applying a gate voltage to the system. 
Thus motivated, we change the on-site energy of the molecular orbital from the symmetric
value, noticing that the particle-hole symmetry is removed for 
$\epsilon_d \neq -U/2$. Therefore, we cannot expect the positive 
and negative half of the spectrum to be symmetric. In Fig.\ref{fig:assymetric},
we plot the energies and spectral weights of the YSR bound states
both for positive and negative energies. Here, the Coulomb interaction is fixed at $U=0.10$
while varying the on-site energy of the orbital for both ferromagnetic 
Fig.\ref{fig:assymetric} (a) and anti-ferromagnetic Fig.\ref{fig:assymetric} (b) exchange interactions.
It can be noticed, that for both ferromagnetic and anti-ferromagnetic exchange interactions, the molecular ground state remains in the doublet regime for small negative on-site energy, 
which leads to the emergence of two pairs of YSR states. While the YSR states in 
the doublet phase coincide with the edges of the superconducting gap at $\epsilon/U=0$,
they are shifted inside the gap with increasing $-\epsilon_d/U$, and eventually transition into
the singlet phase where only a single pair of YSR states with finite weight exists. We have chosen Coulomb
interaction in such a way that at  the symmetric points ($\epsilon_d = -U/2$) we are in the singlet phase. 
With further increase of $-\epsilon_d/U$, the system re-enters the doublet phase where two YSR states 
re-appear.
This re-entrance of the phases can be attributed to the nonlinear behavior of the
doublet ground state energy as a function of on-site energy and can be understood 
in terms of the large $\Delta_\text{sc}$ effective model. The energies of effective
model, Eq. (\ref{eqn:effhal}), in the absence of the exchange interaction are $\epsilon_d$ and
$\epsilon_d + U/2 \pm\sqrt{\tilde{\Gamma}^2 + (\epsilon_d + U/2)^2} $. While the former 
energy refers to the singlet state, depending linearly on $\epsilon_d$, the latter
energies refer to the doublet states, depending nonlinearly on $\epsilon_d$. From
this observation, it is evident that the system is in the doublet phase for small $\epsilon_d$ and in the singlet phase for intermediate values of $\epsilon_d$. It is also possible to always remain in the doublet phase by varying $U$ and $\tilde{\Gamma}$ such that the ration $U/\tilde\Gamma$ remains nearly unchanged.

\subsection{Effect of the anisotropy field}
\begin{figure}[t]
    \includegraphics[width=\columnwidth]{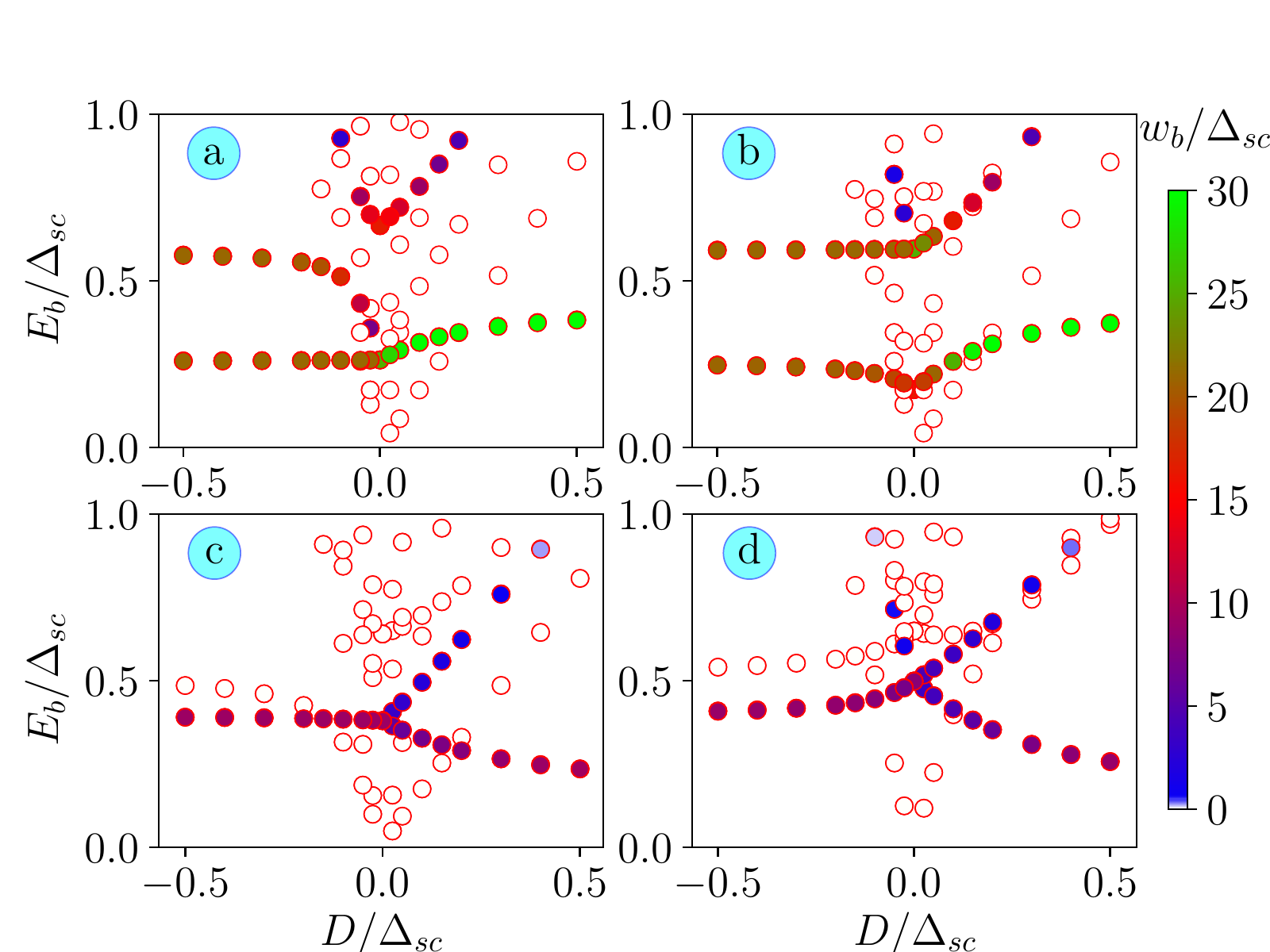}
    \caption{Bound state energies and the color-coded spectral weights  of the molecule  for different value of
    an anisotropic field for $U = 0.07$ (a,b), 0.10 (c,d), $J_\text{ex}/\Delta_\text{sc} = -0.2$ (a,c), 0.2 (b,d).}\label{fig:specJD}
\end{figure}

Large spin molecules are always subject to more or less strong anisotropic fields due to spatial
structure and the intrinsic spin-orbit
coupling of the substrate. In order to describe a physical molecular systems, we
add an uniaxial anisotropy term  to the
core spin Hamiltonian, that is, $H_\text{S} =  DS_z^2 $.
The anisotropy field lifts the degeneracy
of the singlet and doublet states, creating possibilities for the emergence of  
additional YSR states inside the superconducting gap. We also notice that 
positive (negative) values of the parameter $D$ refers to 
uniaxial anisotropies which, respectively, leads to a low (high) spin ground state. 
In Fig.\ref{fig:specJD}, we plot the evolution of the YSR states  for
various combinations  of the Coulomb  and exchange interactions.
We have chosen $U  = 0.07 $ and $0.10$, such that the ground state is in the doublet
and singlet phase, respectively, both for ferromagnetic and anti-ferromagnetic exchange
interactions. In the presence of exchange interaction and anisotropic field 
there many more states appear within the gap. But not all of them have nonzero 
single particle spectral weight. The states with zero spectral weight 
is denoted as empty circles in the figures.

First, consider the lower Coulomb interaction, $U=0.07$ Fig.\ref{fig:specJD} (a), (b). 
For ferromagnetic  interaction, Fig.\ref{fig:specJD} (a), the two YSR states
of  the doublet phase, evolve into three distinct YSR states for small negative
values of the anisotropy field $D$. For larger negative values of anisotropy 
parameter $D$, however, one of these YSR
merges into the edge of the superconducting gap, while the energies of the two 
remaining YSR states remain almost constant.
On the other hand, for small positive values of the anisotropy parameter $D$,
two YSR states appear out of which one of merges into the edge of the 
superconducting gap as $D$ increases. By contrast, for anti-ferromagnetic 
exchange interaction the picture changes, which can be viewed in 
Fig.\ref{fig:specJD} (b) ($U=0.07$ {and } $J_\text{ex} = 0.2$).
At this value of $U$, ground state is  still in the doublet phase 
and in the absence of anisotropic field we see two YSR state. 
The higher energy YSR state split into two YSR state for negative 
values of $D$ and it reamin as a single YSR state for positive
values of $D$. As it was shown in Fig.\ref{fig:specJD} (b), 
one of the YSR state merges into the continuam states above $\Delta_\text{sc}$ 
with increasing  value of $|D|$. However the lower energy YSR state
at $D=0$ does not split and energy of it increases and saturate for larger $|D|$.
As a result, two YSR state
for negative values of $D$ and one YSR state are visible 
in the positive half of the spectrum.

Next we consider the case for larger Coloumb energy at  $U=0.10$ 
in Fig.\ref{fig:specJD} (c), (d). We are now in the singlet phase and only one 
YSR state appear with finite weight at $D=0$. We have shown the evolution of 
YSR states for negative value $J_\text{ex}$ in the Fig.\ref{fig:specJD} (c). 
Here, a single YSR state at $D=0$ remains as a single YSR state 
for negative values of $D$ while it split into two YSR states 
for positive values of $D$. The energy difference between these two increases 
with $D$.
The YSR states of the molecule for positive value of exchange interaction 
 ($J_\text{ex}/\Delta_\text{sc} = 0.20$) is shown in Fig.\ref{fig:specJD} (d).
Here, a single YSR state at $D=0$ splits into two YSR states for both positive
and negative values of the anisotropic field.

To summarize the effect of the anisotropic field,  we see a huge change in 
YSR states. The number of YSR increases due internal spin excitations
of the molecular core spin for small values of the anisotropic field.
With larger values of the anisotropic field some of the YSR state 
energies move to higher energies and end up mixed up with the 
continuum of states above the superconducting gap.
 Depending on the various experimental conditions and spatial structure of the
    molecule and  substrate
    the actual  value  of $U/\Gamma$, $J_\text{ex}$, $D$ can be different. As a result
    the number of visible
YSR state in scanning tunneling microscopy experiment can be different  even for same molecule.

\section{Summary and conclusion}
In conclusion, we have considered the properties of YSR states created from a magnetic molecule absorbed
on the surface of an s-wave superconductor. The molecule  is modeled
as a  single orbital and a core spin, coupled via an exchange interaction.
The competition between the Kondo
effect and superconductivity  determines the nature of the
many-body ground state and the excited states of the molecule-superconductor complex. Depending on the ratio between the energy scales associated with  the Kondo effect and superconductivity, the ground state of the emerging YSR states can be either a singlet or a doublet.
The induced superconducting order parameters on the molecule shows a discontinuity and changes sign at the singlet-doublet transition
point, related to the different ground degeneracies of the singlet and doublet states.
The exchange interaction is a crucial ingredient of our system as it lifts the degeneracy
of the singlet states. As a result, two pairs of  YSR states appear, in general, however, in the singlet
phase only one of the pairs have finite spectral weight at zero temperature. These
results based on NRG simulations are qualitatively consistent with analytical predictions made for
large $\Delta_\text{sc}$ in terms of the effective model given in Eq. (\ref{eqn:effhal}).

We, furthermore, studied the effects of the on-site energy in a set-up out-of the particle-hole
symmetry point ($\epsilon_d = -U/2$). Here, the Coulomb interaction is fixed such that 
the ground state retains the singlet phase with one pair of YSR states at the particle-hole symmetric point. At a critical point, the system undergoes a transition into a doublet state ground state as the on-site single electron energy $\epsilon_d$ is either increased or decreased. One of the YSR
states approaches the Fermi energy at the transition point. This effects is predicted to be measurable in
experiments since the on-site energy can be changed by means of a gate voltage.

Finally, we have investigated the
effects of a uniaxial anisotropy field, acting on the core spin, on the YSR states. Here, both the exchange and Coulomb energies play crucial roles to determine the number of YSR states. Of importance to notice, is that for small negative values of
$D$, see Fig.\ref{fig:specJD} (a), (c), the number of YSR states changes from three to one.
Hence, keeping the values of the exchange interaction, $J_\text{ex}$, and the anisotropy, $
D$, small and negative value, a continuous variation of $U/\Gamma$ should enable observation 
of a change in the number of YSR states. Excitation spectra of MnPc resolved using 
scanning tunneling microscopy\cite{orbital4} also show similar changes in the  properties of the YSR states.
The spectral weights of the individual YSR states, moreover, show discontinuous changes
across the phase transition. Future studies should involve investigations of  finite temperatures and magnetic field effects the state emerging both inside and outside
the gap.

\acknowledgments
We thank Stiftelsen Olle Engkvist Byggm\"astare and Vetenskapsr\aa det for financial 
support. The computations were performed on resources provided by SNIC through
Uppsala Multidisciplinary Center for Advanced Computational Science (UPPMAX) 
under Project SNIC 2019/8-211. We thank Felix von Oppen  and  Katharina J. Franke for useful discussions.

\end{document}